# Nitrogen-doped graphene based triboelectric nanogenerators

*Giuseppina Pace,* Michele Serri, Antonio Esau del Rio Castillo, Alberto Ansaldo, Simone Lauciello, Mirko Prato, Lea Pasquale, Jan Luxa, Vlastimil Mazánek, Zdenek Sofer, Francesco Bonaccorso** *

Dr. Giuseppina Pace
Institute for Microelectronics and Microsystems - National Research Council (IMM-CNR)
Via C. Olivetti 2, 20864, Agrate, Milan, Italy
*E-mail: giuseppina.pace@mdm.imm.cnr.it*

Dr. Michele Serri, Dr. Alberto Ansaldo, Simone Lauciello, Dr. Mirko Prato, Dr. Lea Pasquale
Fondazione Istituto Italiano di Tecnologia (IIT), Via Morego, 30, 16136, Genova, Italy

Dr. Michele Serri current address: XNEXT srl, via Valtorta, 48, 20127, Milan, Italy

Prof. Zdenek Sofer, Dr. Jan Luxa, Dr. Vlastimil Mazánek
Dept. of Inorganic Chemistry, University of Chemistry and Technology Prague
Technická 5, 166 28 Praha 6, Czech Republic

Dr. Francesco Bonaccorso, Dr. Antonio Esau del Rio Castillo
BeDimensional S.p.A, Via Lungotorrente Secca 3d, 16163, Genova, Italy
*Email: f.bonaccorso@bedimensional.it*

## Abstract

Harvesting all sources of available clean energy is an essential strategy to contribute to healing current dependence on non-sustainable energy sources. Recently, triboelectric nanogenerators (TENGs) have gained visibility as new mechanical energy harvester offering a valid alternative to batteries, being particularly suitable for portable devices.

Here, the increased capacitance of a few-layer graphene-based electrode is obtained by incorporating nitrogen-doped graphene (N-graphene), enabling a 3-fold enhancement in TENGs' power output. The dependence of TENGs' performance on the electronic properties of different N-graphene types, varying in the doping concentration and in the relative content of N-pyridinic and N-graphitic sites is investigated. These sites have different electron affinities, and synergistically contribute to the variation of the capacitive and resistive properties of N-graphene and consequently, TENG performance.

It is demonstrated that the power enhancement of the TENG occurs when the N-graphene, an n-semiconductor, is interfaced between the positive triboelectric material and the electrode, while a deterioration of the electrical performance is observed when it is placed at the interface with the negative triboelectric material. This behavior is explained in terms of the dependence of N-graphene quantum capacitance on the electrode chemical potential which shifts according to the opposite polarization induced at the two electrodes upon triboelectrification.

## 1. Introduction

In the last decade, a new technology, based on the triboelectric nanogenerators (TENGs),[1,2] has raised in the field of energy harvesting, allowing to convert the green mechanical energy available in the environment into electrical power. At the base of TENG working mechanism is the electron transfer occurring upon friction between an electropositive and an electronegative triboelectric material (tribomaterial), which determines the establishment of a triboelectrification field. In a TENG each material is interfaced with an electrode and the two are connected through and external circuit. Due to the induction field generated at the electrodes caused by the triboelectrification field, a current flow through the external circuit, the Maxwell displacement current.[3,4] TENGs allow to efficiently convert the mechanical energy originated from the human body,[5] wind[6] and sea waves motion,[7] into clean and sustainable electrical power.[2,8] Due to the light weight, low cost and sustainable nature of materials employed in TENG, they represents a valid alternative to current mechanical harvesters,[4] providing high efficiency even at low frequency.[9–12]

Strategies implemented so far for improving TENGs power output, aim at increasing either the tribomaterial dielectric response[13] and/or the density of the accumulated triboelectric charges.[14] The introduction of high dielectric constant[15–18] and ferroelectric layers[19–21] at the interface between the tribomaterial and the electrode, has been proved to increase the TENG performance. This is explained by a combination of an enhanced electrode polarization due to an increased dielectric response of the tribomaterial, and a charge blocking effect, which reduces the quenching of the triboelectric charges by impeding their drift from the tribomaterial to the underneath electrode.[14,22–24]

The triboelectrification charge density can be effectively enhanced by increasing the surface area, with the use of nanostructured[25] and nanopatterned[26] tribomaterials. Another promising approach is the introduction of charge trapping additives[27–29] and charge-transporting layers[30] within the insulating tribomaterial. These electroactive additives allow scavenging the triboelectric charges accumulated at the surface of the tribomaterial, by creating charge storage sites more deeply embedded in the triboelectric layer.[14]

2D materials are between the most effective electroactive additives, that are embedded in TENGs either as triboelectric layers[31] or in the form of polymeric composite in the tribomaterial.[32] However, it is still not clear which specific properties make them good electroactive additives in TENGs. Though many materials fall under the same umbrella of 2D materials, their properties are different: h-BN is an insulator,[33] black phosphorous[34] and reduced graphene oxide (RGO)[35,36] are mostly conductive, while $MoS_2$[20,28,37–39] is an intrinsic semiconductor[40] and MXenes[41,42] can be tuned by chemical design over a large range of electrical properties. It is therefore still difficult to identify which are the intrinsic properties of different 2D-materials contributing to the enhancement of the TENGs performances, making it hard to design and select more efficient materials.

We recently demonstrated that the selection of electrodes with a large difference in work function can add a built-in potential increasing the TENG open-circuit voltage.[43] We also anticipated that few-layers graphene (FLG)[44,45] electrodes further boost TENGs performance with respect to highly conductive electrodes, such as gold, thanks to their high surface area.

While other works have addressed the role of polymeric interfacial layers, our aim here is to assess the role played by the chemical and electronic properties of 2D additive in affecting the TENGs performance. In particular, we provide evidence of the role played by doped graphene[46–48] in rising the electrode capacitance and therefore in boosting the TENG performance. We specifically selected nitrogen-doped graphene (N-graphene) since it owns higher capacitance than undoped graphene.[49–54] Its introduction as an interfacial layer between the electrode and the triboelectric material, allowed a 3-fold increase in TENGs' power output.

As previously shown in the field of capacitors[49,52,54,55] and batteries,[50,51] the electrical and capacitive properties of N-graphenes are correlated not only to the dopant concentration, but are also strictly dependent on the relative presence of graphitic and pyridinic nitrogen sites (N-graphitic, $N_{graph}$, and N-pyridinic, $N_{pyr}$, respectively).[56–61] We here compare TENGs based on N-graphenes with low and high dopant concentration, as well as samples with different $N_{pyr}$ and $N_{graph}$ content ratio. We investigate the semiconductive and capacitive properties of those N-graphenes by Raman and impedance spectroscopies, providing a direct correlation of these properties with the TENGs performance. The synergistic effect between $N_{pyr}$ and $N_{graph}$ sites is shown to be correlated with the capacitance and resistance ratio (C/R), being directly reflected in the TENGs' power output.

The above observations are explained also in terms of the influence of the quantum capacitance (defined as the variation of the charge density $dn$, with respect to the variation of the chemical potential $d\mu$, $C_q \propto dn/d\mu$) on the electrode capacitance. In fact, the N-doping does not only change the charge mobility by donating additional mobile electrons to the graphene lattice and opening up a band-gap at the Dirac point,[46,62] but also strongly modify the graphene chemical potential therefore its quantum capacitance,[63–66] which is also affected by the presence of different N-sites.

Our findings highlight the fundamental role played by the electrode capacitance in boosting TENGs performance, opening the way to new design principles to improve both TENGs structure and material components.

## 2. Results and Discussion

### 2.1. Material Synthesis and Characterization

*2.1.1. N-doped graphene synthesis*

Nitrogen doped graphene identified as N1 and N2, were prepared from thermal reduction of graphite oxide in reactive ammonia atmosphere at 800°C. Sample N1 was obtained from thermal reduction of graphite oxide prepared by Hofmann methods,[67] which is typically rich in hydroxyls and epoxide functional groups. Sample N2 was prepared by thermal reduction of graphite oxide prepared by Tour method.[67] The thermally reduced graphite oxide denominated as r-N was prepared by reduction of Hofmann graphite oxide in nitrogen atmosphere at 800°C. The reductive processing under $N_2$ is typically responsible for low doping and favors the formation of graphitic nitrogen.[67]

*2.1.2. N-doped graphene Characterization*

*X-ray Photoelectron Spectroscopy (XPS).* Wide X-ray photoelectron spectra (XPS) allow to determine the relative content of the different elements present in the materials under study (see also supporting information, SI). The nitrogen content was found to be 4.0, 7.3 and 0.9 at %, for N1, N2 and r-N respectively, consistently with the trend observed in the combustion analysis (SI). N-graphenes are typically characterized by different nitrogen valence states due to the presence of C-N polymorphs that are convoluted in the N 1s region of the XPS spectra (390-410 eV). In this region, the presence of pyrrolic, pyridinic ($N_{pyr}$) and graphitic ($N_{graph}$) nitrogen, as well as nitrile, aminic and amidic groups can be detected (see table in SI).[68] The contribution of each N-functional group found in the N 1s binding energy region per each N-graphenes is shown in **Figure 1**. The N 1s peak convolution is shown to be composed of graphitic-N ($\sim$ 401.1 eV), pyridinic-N ($\sim$ 398.0 eV), pyrrolic-N (400.1 eV), aminic-N ($\sim$ 399.3 eV) and oxidized-N ($\sim$ 402.5 eV). **Figure**

**1c** confirms that the $N_2$ reduction process used for the preparation of the r-N sample, determine the low N-doping concentration and the favourable formation of $N_{graph}$ and $N_{pyr}$.

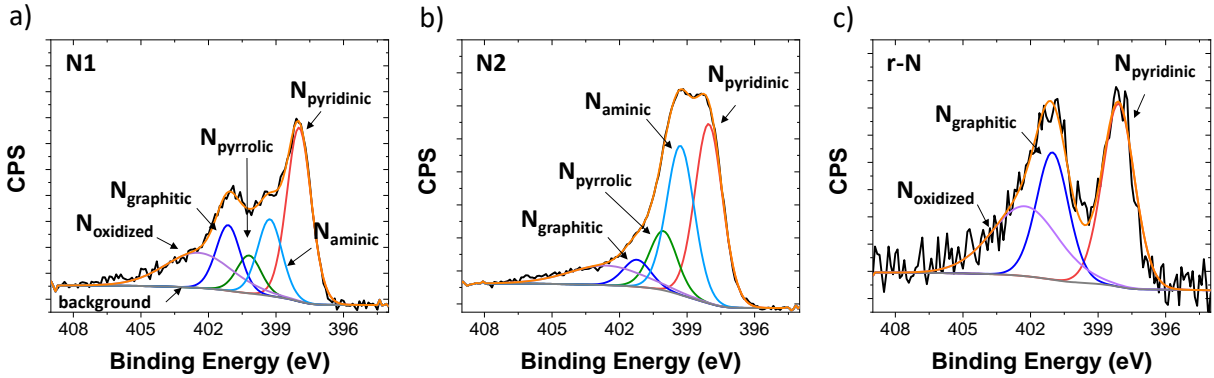

**Figure 1:** N 1s XPS spectra for N-doped graphene materials: a) N1; b) N2; c) R-N.

Oxygen content can also contribute to the capacitance of the N-graphenes.[69] The material with the lowest amount of oxygen is r-N (1.3 at% in oxygen), whereas N1 and N2 have an oxygen content of 2.7 and 3.7 at % respectively. The low oxygen content found for all materials demonstrates the efficient removal of oxygen upon thermal reduction with the successful introduction of nitrogen into the graphene lattice. We also acquired and compared high-resolution core C 1s spectra (SI) for all N-graphenes in which the different contributions of C-$sp^3$, C-$sp^2$ and C-heteroatom bonds can be found.

*X-ray diffraction (XRD).* X-ray diffraction (XRD) spectra are reported in the **SI.** XRD patterns are dominated by two broad peaks centered at ~ 27° and 43° consistently with the defected and amorphous structure of N-doped graphenes.[67]

*Scanning Electron Microscopy (SEM).* Scanning Electron Microscopy (SEM) images of the N-graphenes under investigation can be found in the SI. All samples have shown typical exfoliated structures with wrinkles and highly corrugated surface as also shown in previous work.[33] The

X-ray elemental maps show the homogeneous distribution of each compositional element all over the mapping area (EDX elemental maps in SI).

*Raman.* The typical Raman spectra of N-doped graphenes in the range between 1000 and 3000 cm$^{-1}$,[57,70–72] resemble that of defective graphene as shown in **Figure 2**.[73,74] At approximately 1580 cm$^{-1}$ the G peak can be observed, associated with the in-plane vibrations of sp$^2$-bonded C-atoms (E$_{2g}$ phonon at the Brillouin zone center).[75] At ~1350 cm$^{-1}$ we find the D peak originated by the breaking of the symmetry of the sp$^2$ carbon rings (A$_{1g}$ breathing mode) and typically observed in the presence of defects such as edges, sp$^3$ hybridized carbons, and vacancies.[76,77] The presence of the D´ band at ~ 1615-1620 cm$^{-1}$, is also associated with the disorder-induced phonon mode due to crystals defects.[62,78,79] The in-plane nitrogen heteroatom substitution of sp$^2$ carbon atoms is also a type of defect which activates the D and D´ bands, as it induces compressive/tensile strain on the C-C bonds. The 2D peak found at ~ 2700 cm$^{-1}$ **(Figure 2b)** is an overtone of the disorder-induced D-band assigned to an intervalley double resonance scattering of two phonons around the K-point of the Brillouin zone.[77,78] Typically the 2D band reduces its intensity and is replaced by a bump in defective graphene as is the case of N-graphenes. As anticipated by XPS analysis, a variable amount of different C and N bonding structure and defects appears upon N-doping. These bonds variety causes the large broadening of the D and G peaks region,[76] due to the presence of supplementary peaks which need to be deconvoluted before an accurate analysis can be carried out, as already shown for N-graphene[80–82] and graphene oxides.[83] According to previous work, two extra peaks are typically observed when fitting the D, G and D´ peaks' region (1000-1800 cm$^{-1}$),[77] a small peak located between 1150 and 1200 cm$^{-1}$ (D*)[84] and a broad peak between 1500 and 1550 cm$^{-1}$ (D″).[83] Their intensity, position and linewidth are correlated with the change of hybridization occurring in the graphene lattice in the proximity of the defect C atom.[74,77] The D″ band is typically observed in presence of $sp^3$ carbon as in C-OH, C-NH, C-H, and is related to amorphous phases since its intensity decreases with the increase of crystallinity.[83,85] Other multiple peaks contributing to this region have also been reported in studies performed on defective graphene monolayers.[83,84,86] Our analysis performed over ~100 spectra acquired per each sample, confirms that the best fitting is obtained if an additional peak located at ~ 1500-1550 $cm^{-1}$ is also considered. The D* peak is not identifiable, probably due to its low intensity compared to the broad D peak, and a single Lorentzian peak can fit the D peak region (**Figure 2a and SI**). The fit of the region between 2000 $cm^{-1}$ and 3400 $cm^{-1}$ highlights the presence of three broad peaks, which have been assigned to the 2D (~ 2700 $cm^{-1}$), the D+D′ (~ 2940 $cm^{-1}$) and the 2D′ (~ 3175 $cm^{-1}$) bands (**Figure 2b and SI**).[87] Typically, the ratio between the D and G peaks intensities ($I_D/I_G$), is associated to the degree of structural defects in graphene derivatives.[76,78,88,89] As expected, the D-band is strongly influenced by the presence of doping induced defects and $I_D/I_G$ is higher than 1 for all samples (**Figure in SI**).

According to previous reports[90–93] the presence of electronic doping in graphene is identified in the Raman spectra through the observation of i) the shift to higher energy of the G peak position (stiffening) for both p- and n-doping, ii) the decrease of the G peak full width at half maximum, FWHM G, found in p- and n- doping iii) the shift of the 2D peak position to lower energies for n-doping, and to higher energy for p-doping iv) the decrease in the ratio between the 2D and G peak intensities $I_{2D}/I_G$, with increasing doping.

Therefore, the analysis of the deconvoluted Raman spectra (**Figure 2**) allows to discriminate between doping processing that produced N-graphenes types with more structural defects, and samples that retained the π-conjugation undergoing a better electron doping, *i.e.*, samples with improved semiconductivity properties.

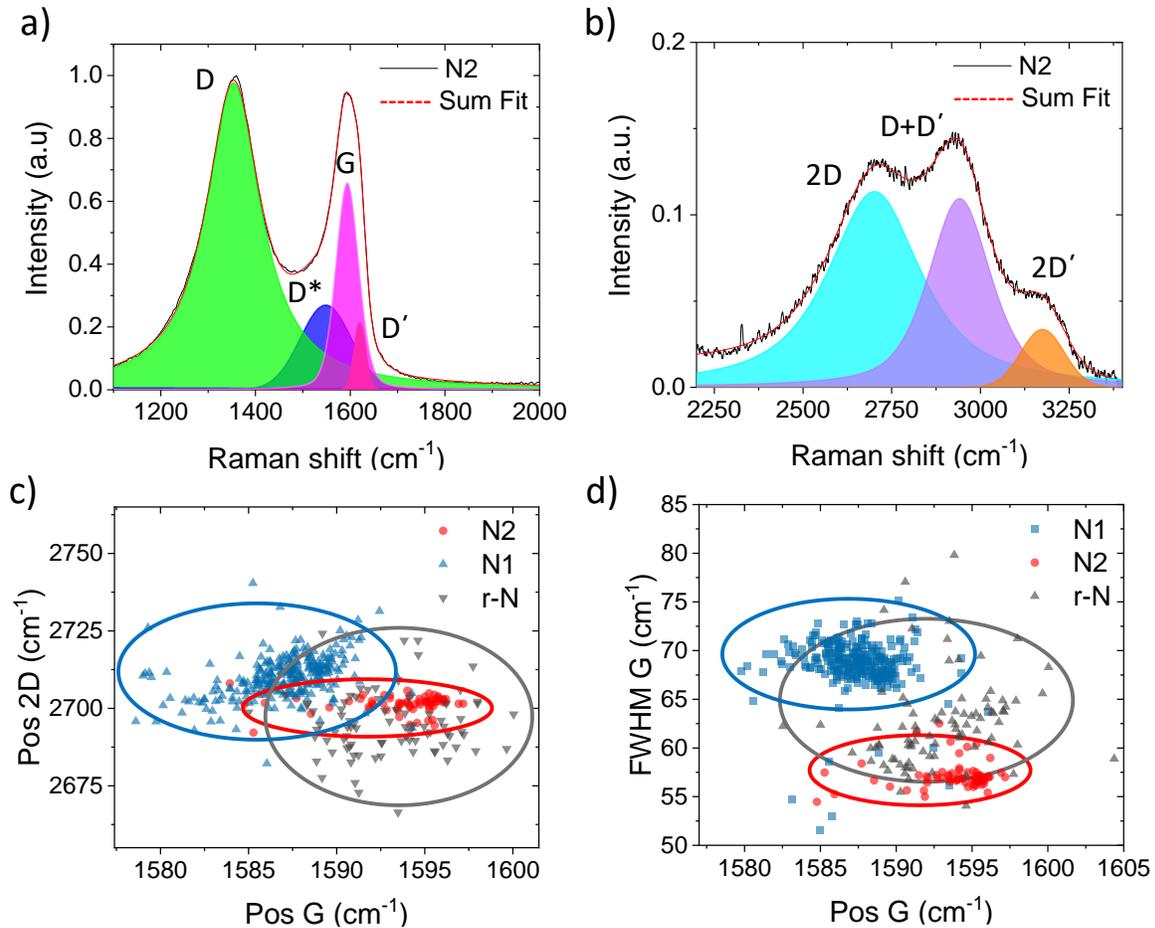

**Figure 2:** Raman spectra analysis. Per each sample at least 100 spectra have been acquired and analyzed. a) N2 single spectra fitting in the D and G peaks region showing the presence of the D, D*, G and D′ peaks; b) N2 single spectra fitting in the 2D region showing the presence of the 2D, D+D′ and 2D′ peaks; c) Distribution of the 2D peak position versus the corresponding G peak position; d) Distribution of the FWHM of the G peak (FWHM G) versus the corresponding G peak position. Continuous lines in c) and d) are guides for the eyes to identify the region of higher data density per each N-graphene sample.

**Figure 2d** shows that all N-graphenes present a blue shift of the G peak compared to graphene where it is typically found at ~ 1585 cm$^{-1}$.[84] However N2 provides a stronger blue shift compared to N1 and r-N, with a peak centered at 1595 cm$^{-1}$ as expected for impurity-doped graphene[91,92] and electrostatically doped graphene.[88,90,94] The G peak distribution centers at 1591 cm$^{-1}$ for r-N and at 1587 cm$^{-1}$ for N1.

The lower energy value of the G peak found for N1 (1587 cm$^{-1}$) compared to N2 (1595 cm$^{-1}$) and r-N (1595 cm$^{-1}$), and the higher energy found for the 2D peak position (2712 cm$^{-1}$ in N1, 2702 cm$^{-1}$ in N2, 2695 cm$^{-1}$ for r-N), allow us to predict that N1 is a more defective material than N2 and r-N samples.[95] Instead, we notice a sharper G (from 1585 to 1597 cm$^{-1}$) and 2D peaks (from 2688 to 2710 cm$^{-1}$) distribution (**Figure 2c**), and a smaller FWHM (G) for N2 (**Figure 2d**, from 55 to 60 cm$^{-1}$), suggesting a less defective material and better electronic doping compared to N1 and r-N.[90,95]

The lower energy found for the G and D peak position in r-N compared to N1 is consistent with its low doping level due to the reductive processing in N$_2$. In previous reports, the I(2D)/I(G) ratio was found between 0.1 and 0.2 for all samples,[91] with a value closer to 0.2 for the stronger Fermi Level (E$_F$) shift occurring in graphene doping induced by impurities.[88,94] A value I(2D)/I(G) closer to 0.2 found for N2, along with the stronger blue shift of the G band and its smaller FWHM, further confirms the better electron doping level obtained in N2 with respect to N1 and r-N.[96]

The ensemble of the data and analysis presented above allows highlighting the chemical and electronic properties of the three N-graphenes useful for the following discussion. The N$_2$ reduction treatment undergone by r-N determines a low N-content which is accompanied by the formation of a less defective sample whit the prevailing presence of N$_{graph}$ and N$_{pyr}$ sites. The Raman spectra confirm the presence of electronic doping for this sample associated to the presence of N$_{graph}$ sites. The N1 sample is more defective as expected by the Hoffman method used for the graphene oxide preparation,[67] and the higher N-doping compared to r-N is not accompanied by an effective electronic doping as shown by the Raman spectra. The N2 sample is characterized by a higher amount of N-doping compared to the other samples. Nevertheless, its electronic doping is strongly highlighted in the Raman spectra.

## 2.2. TENGs Fabrication and Characterization

In the simplest TENG's configuration, the contact-separation mode, two conductive electrodes are separated by an air gap with a distance usually ranging from a few mm up to 1 cm. Each electrode is covered with an insulating triboelectric material to form a triboelectrode and an external circuit ensures the electrical connection between the two electrodes. Under an external mechanical input, repeated contact and separation cycles occur between the two tribomaterials, determining their electrostatic charging. Due to the opposite charges which accumulate at each tribomaterial, an electrostatic field is generated between the two triboelectrodes and an induction current flows through the external circuit. Such triboelectrification field determines the flow of an induction current, the Maxwell displacement current, that is responsible for the power generated by TENGs.[4]

In this work, TENGs operating in contact separation mode have been fabricated using FLG-based flexible electrodes prepared following the previously reported procedure.[43] Briefly, a FLG dispersion[97–99] was used as a precursor for the preparation of a slurry, which was deposited by doctor blade on EVA (ethyl-vinyl acetate) flexible substrates. As tribomaterials, we selected commercial nylon and polyvinylfluoride (PVDF) membranes. This choice is based on their relative position along the triboelectric series, specifically nylon is a positive tribomaterials and PVDF is a negative one.[100] The basic TENG device structure contains a FLG electrode/PVDF and a FLG electrode/nylon triboelectrode (FLG/PVDF /air/ nylon/FLG) (**Figure 3**). Being the positive (nylon) and negative (PVDF) triboelectric materials invariant across the different tested TENGs, there is no variation of the triboelectric layer surface potential.

The N-graphenes have been introduced as an interlayer between the tribomaterial and the FLG-based electrode. The deposition of the different N-graphenes from solution dispersions in isopropanol or ethanol did not provide reproducible TENGs power output, likely due to the diffusion and aggregation of the N-doped graphene within the flexible and soft FLG electrode during the device operation. We therefore, prepared a dispersion of the three samples of N-doped graphene into a polyurethane (PU) glue. PU is well known to form stable composites with graphene and graphene derivatives.[101,102] This strategy added multiple benefits to the device's performance. Firstly, upon drying, the N-graphene functionalized PU glue formed a uniform and stable interface between the FLG electrode and the tribomaterial (~ 5 μm thick, SEM images **Figure 3b and 3c**). Secondly, the diffusion of the N-doped graphene was impeded by the solid-state form of the dried glue. Thirdly, the PU film introduced a polarizable dielectric interlayer between the electrode and the tribomaterial.

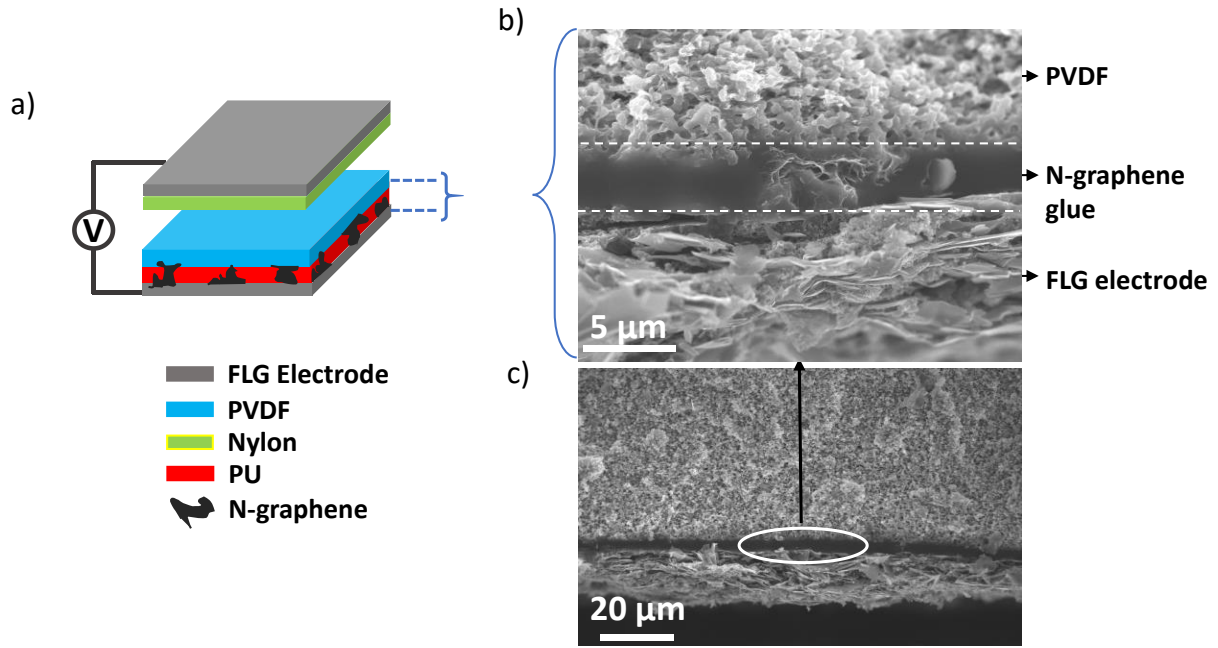

**Figure 3:** a) Schematic of the TENG structure used in the contact-separation operation mode. b) and c) show the SEM cross-section images of the trioelectrode embedding the PVDF membrane.

The N-graphene functionalized PU layer was inserted either at the graphene/nylon interface, *i.e.*, at the interface with the positive tribomaterial (FLG/PVDF // Nylon/PU-X/FLG, **Figure 4a**) or at the FLG/PVDF interface, *i.e.*, at the interface with the negative tribomaterial (FLG/PU-X /PVDF // Nylon/FLG, **Figure 4b**). Both **Figure 4a** and **4b** show the instant power output of TENGs fabricated with no interfacial layer; with only the PU interlayer and with each of the three different N-graphene functionalized PU glues (PU-N1, PU-N2, PU-r-N).

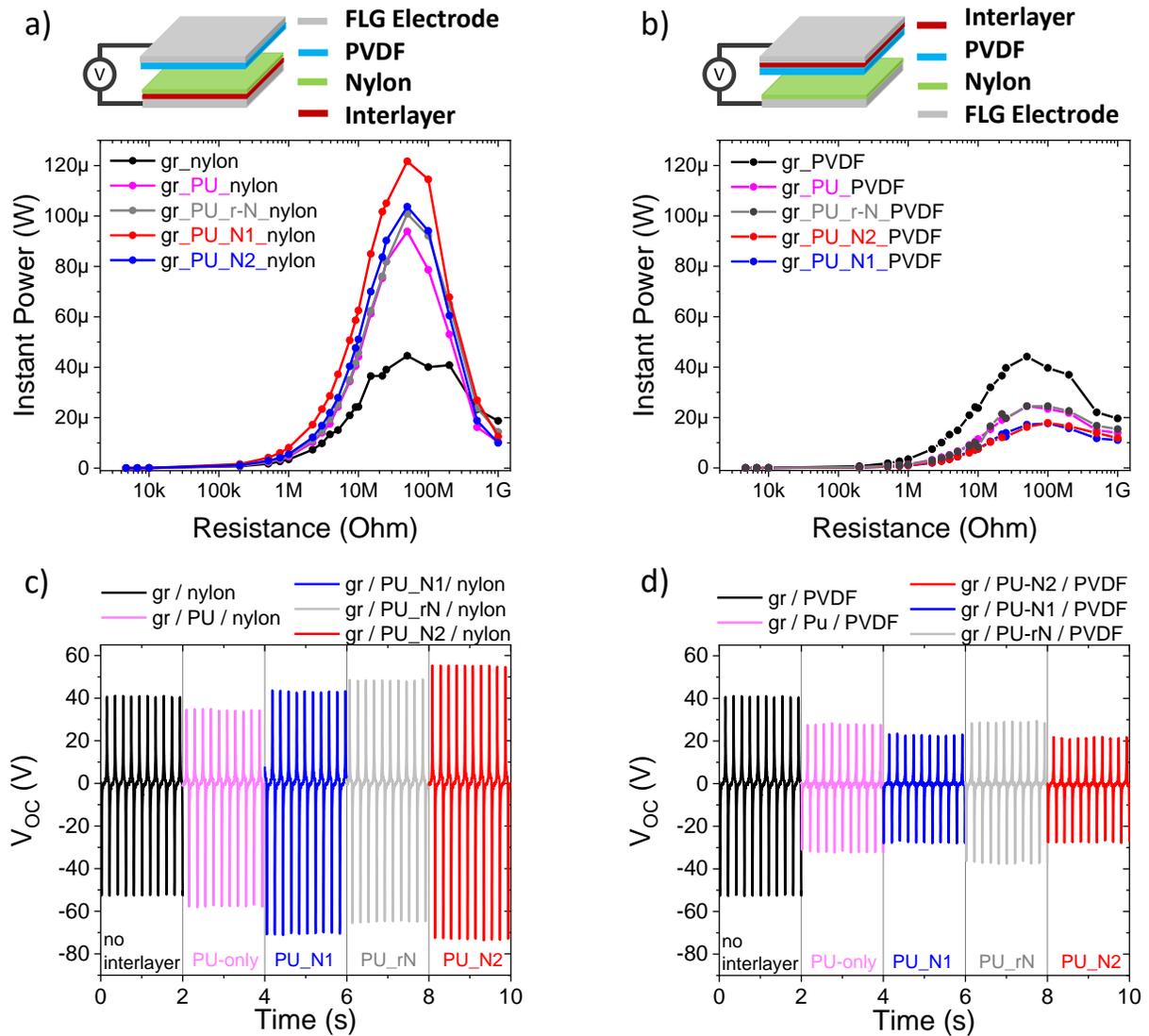

**Figure 4:** TENGs instant power output measured for two device configurations: a) and c) the functional Pu interlayer is introduced at the interface with the positive tribomaterial (nylon) (FLG/PU_X/nylon // PVDF/FLG); RMS power: 6.95 μW, PU_N1; 8.25 μW, PU_N2; 6.45 μW, PU_r-N; 6.3 μW, PU-only; 5 μW for TENGs with no interlayer. b) and d) the interlayer is introduced at the interface with the negative tribomaterial (PVDF) (FLG/nylon // PVDF/PU_X/FLG). c) and d) Open Circuit Voltage (V$_{OC}$) measured after 200 s of continuous cycling where V$_{OC}$ saturation conditions are reached. (Applied force 10 N, operational frequency 5 Hz, 4×4 cm$^2$, 5 mm; V$_{OC}$ acquired with 40 ΩM probe).

The addition of the different N-graphene functionalized PU interlayers at the gr/nylon interface (**Figure 4a**) determines a rise in TENG power output for all types of N-graphene (104 µW for N1, 122 µW for N2, 101 µW for r-N), and the beneficial effect of the dielectric polarizability of the PU is observed in PU-only devices (94 µW). A 3-fold increase in power output is achieved with N2 compared to neat FLG electrode (44 µW, gr/nylon), corresponding to ~ 170 % increment, while ~ 30 % rise in power output is achieved when compared to TENGs embedding only the PU glue (FLG/PU/nylon).

It is straightforward to notice that when the N-graphene functionalized PU interlayer is inserted at the FLG/PDVF interface (**Figure 4b**, FLG/PU-X/PDVF) a decrease in devices' performance is observed. A possible explanation for such different behavior can originate from the different built-in potential generated by a difference in electrode work function, therefore in its surface potential, associated to the presence of the different N-graphenes on top of the FLG. To ascertain this point, we performed Ultraviolet Photoelectron Spectroscopy (UPS) measurement (**Figure in SI**). UPS shows that there is no difference in work function, between the different graphene electrodes topped with each N-graphene. Therefore, the difference in TENGs electrical output in **Figure 4a** and **4b** cannot be explained with a difference in surface potential and therefore with a difference in TENG built-in potential.

As described in the introduction, previous work already reported the improved performance obtained upon addition of 2D materials as charge trapping additives.[103] However, this mechanism implies that the electrostatic charges harvested by the tribomaterial should migrate from the top contact surface to the 2D material trapping sites. There are two main shortcomings for this interpretation: i) our tribomaterials are 200 µm thick and charge transfer through an insulator cannot occur over such long-distance;[104] ii) since our N-graphenes are intimately connected to the underneath FLG electrode, as shown in **Figure 3**, such process would determine the leakage of the triboelectrification charges towards the electrode, decreasing the TENGs' performance, which is the opposite of what we observe.

To verify the role played by the N-graphene in our TENGs, the impedance response of the PU-only and the N-graphene functionalized PU interlayers (PU_N1, PU_N2, PU_r-N) sandwiched between two gold electrodes (**Figure 5**) was measured.

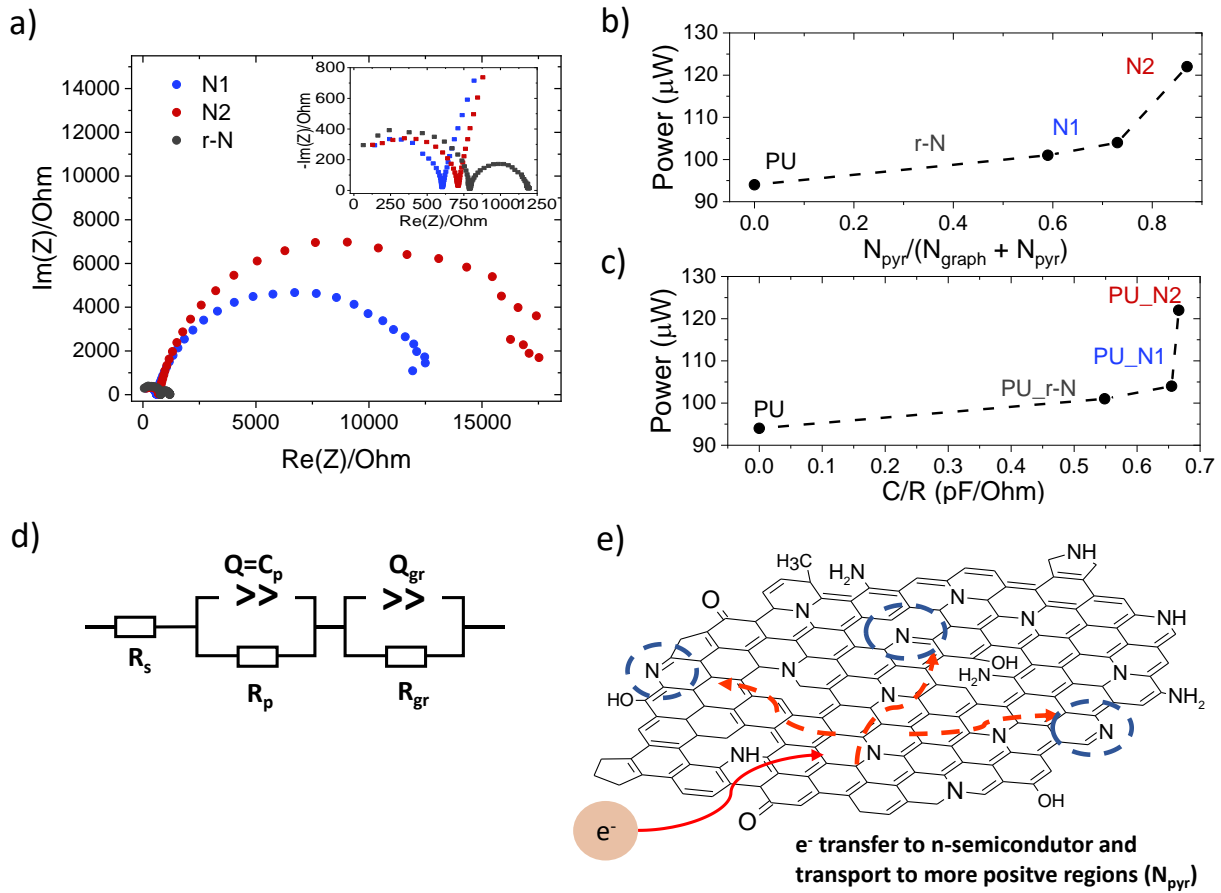

**Figure 5:** a) Impedance spectra acquired on sandwich-like devices composed of Au/PU-interlayer/Au (PU_N1; PU_N2; PU_r-N). The inset shows the impedance spectra in the high frequency region. b) Dependence of the instant peak power from the $N_{pyr}/(N_{graph} + N_{pyr})$ ratio. c) Dependence of the instant peak power from the capacitance vs resistance ratio (C/R). d) Equivalent circuit used to fit the impedance spectra in a). e) Scheme representing the synergistic role of $N_{pyr}$ and $N_{graph}$ in increasing the N-graphene capacitance.

A purely capacitive behavior is observed for the PU-only layer as expected for an insulating dielectric film (**SI**). All the N-graphenes functionalized PU films instead, have shown a semicircle in the high frequencies' region (above 100 Hz), in which electronic, atomic, dipole and ionic relaxation phenomena mostly occur,[105] accompanied by a second semicircle found in the lower frequencies range (below 100 Hz), which is very well defined for the r-N functionalized glue (inset **Figure 5a**). The impedance spectra were modeled with an equivalent circuit containing a resistance ($R_s$, resistance of electrodes and wiring, $R_s \sim 0 \, \Omega$ for all samples) and two RC circuits (**Figure 5d**). A constant phase element (Q and $Q_{gr}$) has been considered per each RC circuit to account for any non-idealities present in the system. The RC component prevailing in the high-frequency range (first semicircle), shows that the constant phase element fitting parameter was equal to 1 for all the layers demonstrating the presence of an ideal polarization capacitance ($Q = C_p$) and an electrode polarization resistance ($R_p$). The impedance data extracted from the equivalent circuits fitting in the high-frequency range are summarized in **table 1**. The capacitance value found for all N-graphene (385 pF, N1; 473 pF, N2; 430 pF, r-N) is roughly a factor two higher than the capacitance of the pristine PU-only layer (210 pF), confirming the favorable effect of N-graphene on increasing the electrode capacitance. This rise in electrode capacitance in presence of the N-graphene functionalized glue is therefore at the base of the improved TENG performance.

**Table 1.** Impedance data from the equivalent circuits fitting in the high-frequency range

| Interlayer | Instant Power (µW) [a] | Capacitance ($C_p$) (pF) [a] | Resistance ($R_p$) (Ohm) [a] |
|---|---|---|---|
| **No interlayer** | 45 | | |
| **PU** | 94 | 210 | |
| **PU_r-N** | 101 | 430 | 784 |
| **PU_N1** | 104 | 385 | 588 |
| **PU_N2** | 122 | 473 | 710 |

[a] Instant power at peak maximum. [b] Polarization capacitance ($C_p$) and resistance ($R_p$) were extracted from impedance acquired on Au/interlayer/Au devices (original data and fitting circuits in SI).

The low-frequency range (below 100 Hz) is dominated by space charge relaxation and conduction processes[105] and the presence of the second semicircle in this region is associated with the charging and conductive properties of the N-graphenes ($C_{gr}$, $R_{gr}$). For PU_ N2 and PU_N1, it is still possible to observe the influence of diffusive phenomena, probably arising from impurities or residual solvent, while the well-defined second semicircle found for r-N indicates that this N-graphene can more effectively create conductive channels within the insulating PU film. This phenomenon is consistent with the more extended conjugation of r-N due to the low N-doping, inferring to r-N a higher conductivity compared to N1 and N2.[106] The lower capacitance found for N1 compared to N2 and r-N in the high-frequency range, is also consistent with the Raman analysis. The latter has shown that this material is less electronically doped, likely providing less efficient charging channels along the N-graphene.

Though impedance spectroscopy demonstrates that the improved TENG performance with N-graphene is related to the increased of electrode capacitive charging, the trend observed for the capacitance ($C_{N1}>C_{r-N}>C_{N2}$) does not follow the TENGs' instant power ($P_{N2}> P_{N1}>P_{r-N}$).

To provide a better interpretation of this observation, it is necessary to consider two important factors. Once the triboelectrification field is settled between the two tribomaterials the electrodes polarize with opposite charges, *i.e.*, the Fermi level ($E_F$ or the chemical potential µ) shifts up in energy for negative polarization, and down in energy for positive polarization (**Figure 6**). Since the quantum capacitance ($C_q$) is defined as the variation in charge density (n) according to a variation of μ ($C_q \propto dn/d\mu$),[107,108] and since we are dealing with a n-type semiconductor, only for an increase of the $E_F$ due to the negative electrode polarization, we can effectively take advantage of the capacitive charging of the N-graphene (**Figure 6b**).[65] The high negative potential of the FLG electrode allows the electrons to fill-in the N-graphene conduction band. For positive electrode polarization, *i.e.*, lowering the $E_F$, an eventual charge transfer from the N-graphene to the FLG electrode would result in an electrode depolarization, therefore to a loss in open-circuit voltage (**Figure 6a**). This last point is consistent with the observed reduction in power output found for TENG embedding the N-graphene PU interlayers in the negative triboelectrode (FLG/PU_X/PVDF).

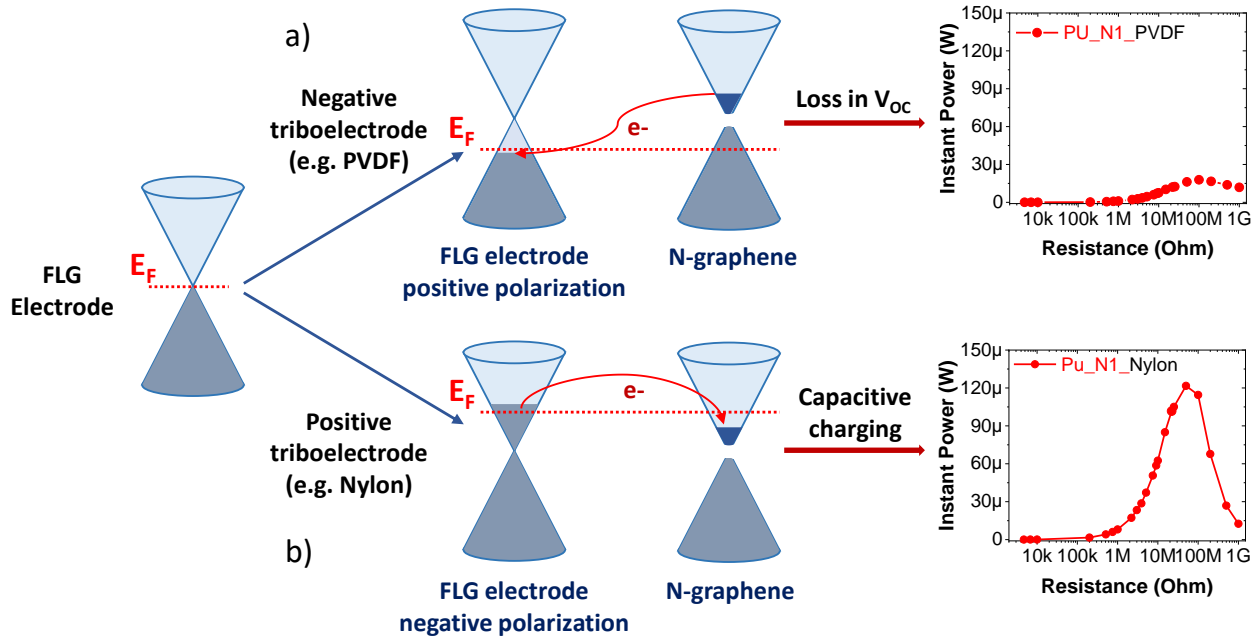

**Figure 6:** Scheme of the mechanism at the base of a) the loss in induction potential ($V_{oc}$) when N-doped graphene is at the interface with negative tribomaterial, b) the improved performance of TENG when N-doped graphene is at the interface with a positive tribomaterial.

Furthermore, it is important to recall that the chemical potential of N-graphenes further depends on the relative content and distribution of $N_{pyr}$ and $N_{graph}$. While all the N-groups have shown electrochemical activity,[56] much attention has been devoted to the main role played by $N_{pyr}$ and $N_{graph}$ in leveraging the electronic properties of N-graphenes.[57,59,61] More recently, the importance of balancing their relative ratio ($N_{pyr}/N_{graph}$), rather than their absolute concentration, has been highlighted as causing the improvement of the electrochemical activity of N-graphene.[58]

The validity of this finding is shown in **Figure 5b** in which the instant power of our TENG nicely follows the $N_{pyr}/N_{tot}$ ratio ($N_{tot} = N_{pyr} + N_{graph}$). Other possible correlations between the TENGs' power and the absolute value of $N_{pyr}$ and $N_{graph}$ were investigated, but no straightforward correlation was observed (**Figure SI**). Importantly, a good correlation between the instant power and the relative ratio between the capacitance and resistance (C/R) of the PU-N-graphene was found (**Figure 5c**).

The relation between the $N_{pyr}/(N_{graph} + N_{pyr})$ and C/R ratio and their variation with the different N-graphenes, can be better understood according to previously reported density functional theory (DFT) calculations and experiments. The concomitant presence of $N_{graph}$ and $N_{pyr}$, generate localized states with opposite electron affinities. The distribution over the graphene layer of those different N-sites determines the formation of puddles of electrons rich and electrons deficient regions, located respectively around $N_{graph}$ and $N_{pyr}$ sites.[109] This creates a spatial variation of the graphene chemical potential, and of the density of stated (DOS) at the $E_F$.[110] The combination of n-type semiconductor properties provided by $N_{graph}$ sites allowing the electrons to flow-in, with the presence of positive potential depth at the $N_{pyr}$ sites permitting to store the injected charges, is therefore the condition to increase the electrode capacitance (see scheme

**Figure 5e**). This perfectly translates with the need for a material that has a large capacitance and low resistance, which is the reason for the observation of the nice correlation of the TENGs power with the C/R and $N_{pyr}$ vs $N_{graph}$ relative content.

## 3. Conclusion

We presented a thorough chemical, structural and electronic investigation of N-doped graphenes and revealed the fundamental parameters needed to be considered for the better design of 2D materials-based TENGs. We demonstrated that an increase in power output is obtained due to the rise of the electrode capacitance associated with the presence of N-doped graphene. Furthermore, the ensemble of our observation points out that proper chemical doping of graphene needs to be coupled with good electronic doping to effectively contribute to the increase of the electrode capacitance. We demonstrated that Raman spectroscopy is a powerful tool to identify the better performing doped-graphene material. The devices' improvement is a function of the relative amount of $N_{pyr}$ and $N_{graph}$ which well correlates to the capacitance/resistance ratio, highlighting the synergistic contribution of both functional groups to the electronic properties of N-graphene. The important role of quantum capacitance in 2D semiconductors, which determines the dependence of the electrode capacitance from its chemical potential, is considered as responsible for the opposite trend in TENGs power output observed when n-doped graphene is inserted as interlayer either at the positive triboelectrode or at the negative one.

This work contributes to understanding the electroactive additive properties, which can lead to rise the power output and helps to lay down consistent and reproducible methodologies for the future design and optimization of TENGs.

## 4. Experimental Section/Methods

*Material Synthesis*.

Nitrogen-doped graphenes and RGO were prepared by thermal reduction of graphene oxide in nitrogen and ammonia atmosphere. As a starting material graphene oxide prepared either by Hofmann or Tour method was used. The detailed synthesis of graphite oxide was reported previously.[111] Graphite with 2-15 µm particle size and 99.9995% purity (Alfa Aesar, Germany) was used for the graphite oxide synthesis.

Samples N1 and N2 were prepared by thermal reduction in an ammonia atmosphere. The sample noted as N1 was prepared from graphite oxide made by the Hofmann method,[67] while sample N2 originate from graphite oxide made by Tour method.[67] Sample r-N was prepared by thermal reduction of graphite oxide made by Hofmann method in a nitrogen atmosphere.[67] For the synthesis, 1g of graphite oxide was placed in a quartz tube with one sealed end connected to the gas control system and vacuum pump. The quartz reactor was first flushed several times with ammonia and nitrogen, respectively. After three-time evacuation and refilling of the quartz reaction tube, the flow of ammonia (and nitrogen for r-N sample) was set on 250 sccm and the tube was inserted in a crucible furnace (100 mm in diameter, 180 mm deep) preheated to 800°C. After 20 minutes the quartz reaction tube was removed from the furnace and the as formed reduced graphene was cooled under the same atmosphere used for the reduction reaction. Finally, the reaction tube was repeatedly flushed with nitrogen, and the sample was removed from quartz tube.

*Materials*. Nylon membranes (0.2 µm pores diameter; 200 µm thick) were supplied by Merk while polyvinylfluoride (PVDF) membranes (0.2 µm pores diameter; 200 µm thick) were provided by GE Healthcare (Amersham). All the other chemicals were purchased from Sigma Aldrich.

*Few-layers graphene (FLG) electrodes.* Few-layer graphene flakes were produced via wet-jet-milling (WJM) as previously reported.[99] The defect-free and high-quality 2D-crystal powder was dispersed to form a homogeneous slurry composed of FLG flakes (WJM), carbon black (Super P, Alfa Aesar) and ethyl vinyl acetate (EVA, Mr Watt Srl) with a weight ratio of 1:250:40, dissolved in a solvent mixture containing butylcarbamate and chlorobenzene (3:7.5 v/v). The slurry was deposited via doctor balding on a clean Al foil, dried under ambient condition at 80 °C overnight. The FLG-based film was transferred onto EVA flexible substrates by hot-pressing the FLG coated Al film on the EVA substrate (80 °C, 15 min). The Al foil was then peeled off the FLG electrode before use. A copper wire connected to the FLG electrode with Ag conductive paste was used to ensure the electrical connection to external loads.

*Polyurethane (PU) functionalized glue* was prepared by mixing the PU glue with the N-graphene dispersion in isopropanol (1 mg/ml of N-graphene in isopropanol, 0.5 mg/ml final concentration). The PU-glue was prepared by dissolving the PU in a solvent mixture containing acetone, toluene, 2-butanone (volume ratio 5:2:2) stirred overnight at 80°. The N-graphene functionalized glue was prepared by mixing equal volumes of the N-graphene dispersion and the PU-glue. The optimal N-graphene concentration was obtained by testing 4 different concentration in a TENG configuration comprising nylon and PVDF as triboelectric materials and aluminum electrodes (data reported in SI). This volume ratio allows to obtain gel-like solutions which provided the proper viscosity to homogeneously cover the FLG and to form a continuous and uniform functionalized glue interlayer (Figure SI). The functional glue was deposited over the FLG electrode via doctor blade and the PVDF (or Nylon) membrane was placed on top of it for gluing. The procedure provides a 5 μm uniform interlayer between the triboelectric membrane and the FLG electrode. The triboelectrode was left drying overnight before its final assembly into the TENG.

*Scanning Electron Microscopy (SEM).* Field-emission scanning electron microscope FE-SEM (JEOL JSM- 7500 FA) was used to acquire the SEM data with the acceleration voltage set to 5 kV. The cross-section SEM images were acquired on a triboelectrode consisting of a flexible FLG electrode interfaced with the PU functionalized glue and PVDF membrane. The triboelectrode was frozen and then broken in a liquid $N_2$ tank before the SEM cross-section images were acquired.

*X-ray diffraction (XRD).* X-ray powder diffraction data were collected at room temperature with a Bruker D8 Discoverer powder diffractometer in the parafocusing Bragg–Brentano geometry. The $CuK_\alpha$ radiation source was employed ($\lambda = 0.15418$ nm, U = 40 kV, I = 40 mA). Data were acquired over the angular range of 10–80° (2θ) and a step size of 0.02° (2θ). Data analysis was performed using the software package EVA.

*Elemental combustion analysis.* Combustible elemental analysis (CHNS-O) was performed using a PE 2400 Series II CHNS/O Analyzer (Perkin Elmer, USA). The instrument was used in CHN operating mode (the most robust and interference-free mode) to convert the sample elements to simple gases ($CO_2$, $H_2O$ and $N_2$). The PE 2400 analyzer automatically performed combustion, reduction, homogenization of product gases, separation and detection. An MX5 microbalance (Mettler Toledo) was used for precise weighing of the samples (1.5–2.5 mg per single sample analysis). Using this procedure, the accuracy of CHN determination is better than 0.30% abs. Internal calibration was performed using N-phenyl urea.

*X-ray photoelectron spectroscopy (XPS).* The XPS measurement was performed with a Kratos Axis UltraDLD spectrometer at a vacuum lower than $10^{-8}$ mbar, using a monochromatic Al Kα source operating at 20 mA and 15 kV and collecting photoelectrons from a $300 \times 700$ μm$^2$ sample area. The charge compensation device was not used. Wide spectra were acquired at pass energy of 160 eV and energy step of 1 eV. High-resolution spectra of O 1s, N 1s, C 1s, and Au 4f peaks were acquired at pass energy of 10 eV and energy step of 0.1 eV. The samples were mounted on the sample holder with copper clips in air. Data analysis was performed with CasaXPS software (version 2.3.19PR1.0). The energy scale was calibrated by setting the Au $4f^{7/2}$ peak at 84.0 eV and C 1s peak (graphitic component) at 284.5 eV.

*Ultraviolet Photoelectron Spectroscopy (UPS).* UPS with He I (hν = 21.21 eV) radiation was carried out after a first XPS analysis was performed on each sample using the same equipment. Spectra were acquired at pass energy of 10 eV and energy step of 25 meV. Photoelectrons were collected from a spot of 55 µm in diameter. A -9.0 V bias was applied to the sample to precisely determine the low kinetic energy cut-off. The energy scale was settled after the binding energy calibration was performed for the XPS measurement.

*The TENGs characterization* was performed employing a home-built set-up comprising a linear motor and load cell. The Open circuit voltage ($V_{OC}$) was measured with an oscilloscope (Tektronix MSO5000) and a voltage probe (40 MΩ, Tektronix). The $V_{OC}$ was measured under continuous contact and separation cycling until no increase in $V_{OC}$ was observed. Current measurements were acquired after 400s of continuous cycling to ensure that TENGs reached the stable maximum $V_{OC}$ necessary to the proper comparison between the different samples. Current measurements were performed with a current amplifier (1211 DL Instruments) and a home-built resistance commutator. The device's active area was fixed at a value of 4 × 4 cm$^2$. The applied force was 10N, the operational frequency was 5 Hz, the maximum separation between the two trioelectrodes was 5 mm. The observed trend was reproducible over different batches and was observed to be stable over a period of months (at least 8 months).

The RMS power was calculated at the instant power peak maximum as a function of the integrated area of full press and release cycle (average peak area over 5 s cycling), corresponding to: 6.95 µW for N1, 8.25 µW for N2 and 6.45 µW for r-N; 6.3 µW for PU-only devices and 5 µW for TENGs with no interlayer ($P_{RMS} = \frac{1}{T}\int_0^T i^2 R dt$, with T being the time interval of a full cycle at 5 Hz).

*Impedance Spectroscopy.* Impedance spectroscopy was performed with a potentiostat/galvanostat (VMP3, Biologic). Gold electrodes were prepared by e-beam deposition of 5 nm Ti and 70 nm Au on polyethylene terephthalate (PEN 50 µm thick, Dupont- Tejin) substrates. Electrical impedance spectroscopy using a 0 V DC voltage ($V_{DC}$) and a10 mV alternating voltage ($V_{AC}$) amplitude was used to study the capacitive and resistive properties of each polyurethane composite glue.

**Nitrogen-doped graphene based triboelectric nanogenerators**

*Giuseppina Pace,*\* *Michele Serri, Antonio Esau del Rio Castillo, Alberto Ansaldo, Simone Lauciello, Mirko Prato, Lea Pasquale, Jan Luxa, Vlastimil Mazánek, Zdenek Sofer, Francesco Bonaccorso*\*

## 1. Chemical and structural characterization of N-graphenes

*a- X-ray Photoelectron Spectroscopy (XPS)*

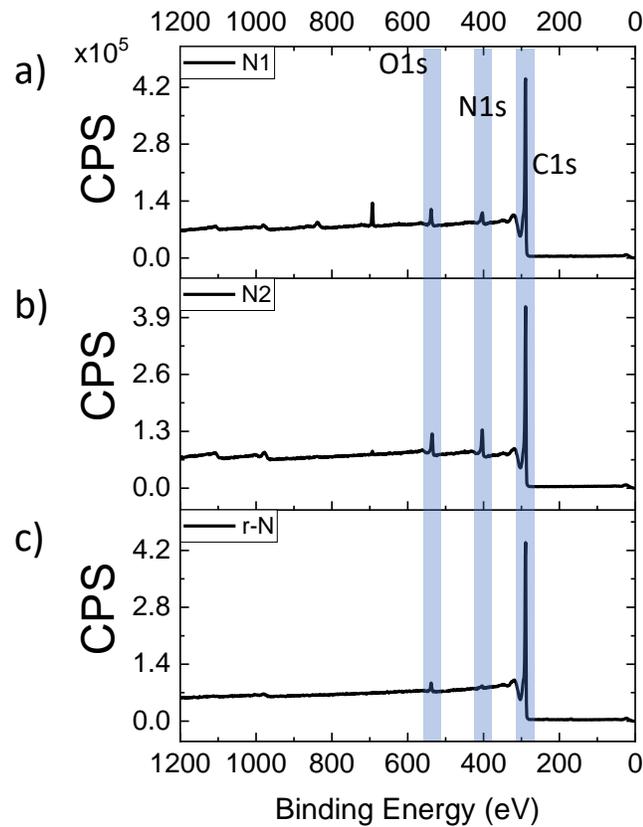

**Figure SI-1:** Wide XPS spectra showing the elemental composition of the N-graphenes investigated in this work: a) N1; b) N2; c) r-N. Fluoride contamination is found at ~700 eV in the N1 sample spectrum (a).

The elemental composition revealed by XPS data analysis is summarized in table SI-1, while the relative content of different C-N components in the N 1s spectra of each sample is reported in table SI-2.

**Table SI-1. Elemental composition derived from XPS**

|     | C 1s (% at) | N 1s (%at) | O 1s (%at) | F 1s (%at) | S 2p (%at) |
| --- | --- | --- | --- | --- | --- |
| N1 | 91.01 | 4.00 | 2.67 | 2.26 | 0.06 |
| N2 | 88.71 | 7.30 | 3.74 | 0.25 | |
| r-N | 97.78 | 0.93 | 1.29 | | |

**Table SI-2. Spectral components in the N 1s XPS spectra.**

| N 1s | Pyridinic N (% at) | Amino N (%at) | Pyrrolic N (%at) | Graphitic N (%at) | Oxidized N (%at) | Total N% |
| --- | --- | --- | --- | --- | --- | --- |
| N1 | 38.66 | 17.06 | 8.40 | 14.54 | 21.33 | 4.00 |
| N2 | 38.44 | 31.91 | 12.66 | 5.78 | 11.21 | 7.30 |
| r-N | 40.20 | | | 27.90 | 31.90 | 0.93 |

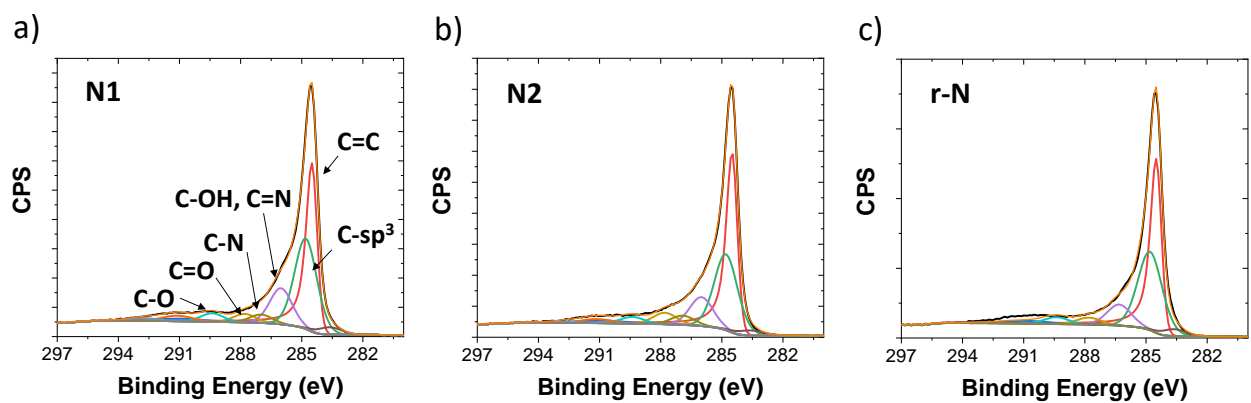

**Figure SI-2:** C 1s XPS spectra for N-doped graphene materials: a) N1; b) N2; c) r-N.

The high-resolution XPS spectra of C1s show the presence of different C valences and C-O bonds with different energy levels: C=C bond at ~ 284.5 eV, C-C bond at ~ 284.8 eV, C-O bond at ~ 286.3 eV, C=O bond at 287.8 eV, O-C=O bond at 289.4 eV and $\pi$-$\pi^*$ interactions at 290 eV.

The peak centered at 284.5 eV, is assigned to the C=C bond and is the major contributor to the C 1s spectra, showing that the delocalized $\pi$-conjugation is preserved upon thermal doping. At lower energies, we find a peak contribution centered at ~ 283.6 eV, assigned to the presence of C lattice vacancies/distortions introduced mainly during the GO reduction process.

The peak around 284.8 eV is attributed to sp$^3$ carbon atoms (C-C) present in all samples but with higher content (35.70 at %) in N1 compared to N2 and r-N, and is related to the presence of structural defects.[1] However due to the presence of adventitious carbon typically found in sample exposed to air contamination, it is not possible to extract from the XPS data information about the relative abundance of C=C bonds vs Csp$^3$ content.[2,3] The presence of the C-O (286.0 eV), O-C=O (289.4 eV) components are associated with residual oxygen functional groups. The peak found at ~291 eV is assigned to $\pi-\pi^*$ satellite structure originated by the extended delocalized electrons,[4,5] and typically observed in large aromatic C structure. In this region (291-293 eV) we find also weak contribution from C-F bonds that arise from the presence of fluorine contamination in N1 and N2 samples.

**Table SI-3. Components in the C 1s XPS spectra.**

| C 1s | C=C including Pi-Pi* (% at) | C-C, C_H, Csp$^3$ (%at) | C-OH, C-O-C, C-N (%at) | C=O (%at) | O-C=O (%at) | C-N (% st) | Defects C 1s (%at) |
|------|------|------|------|------|------|------|------|
| **N1** | 32.74 | 35.70 | 14.24 | 3.33 | 3.33 | 3.28 | 3.11 |
| **N2** | 36.98 | 33.17 | 12.60 | 4.99 | 2.90 | 4.00 | 2.55 |
| **R-N** | 39.47 | 39.38 | 10.90 | 3.42 | 3.37 | | 3.45 |

### b- Combustion analysis

The composition obtained from the combustible elemental analysis shows the different nitrogen concentrations present in each material (1.37 wt.% in R-N, 5.27 wt. % in N1 and 10.9 wt. % in N2).

The concentration of carbon and hydrogen was 86.86 wt. %C and 0.76 wt.% H in sample N1, 77.31 wt. % C and 1.48 wt. % H in N2 and 91.75 wt. % C and 0.78 wt. % H in sample r-N.

### a- X-ray Diffraction Spectra (XRD)

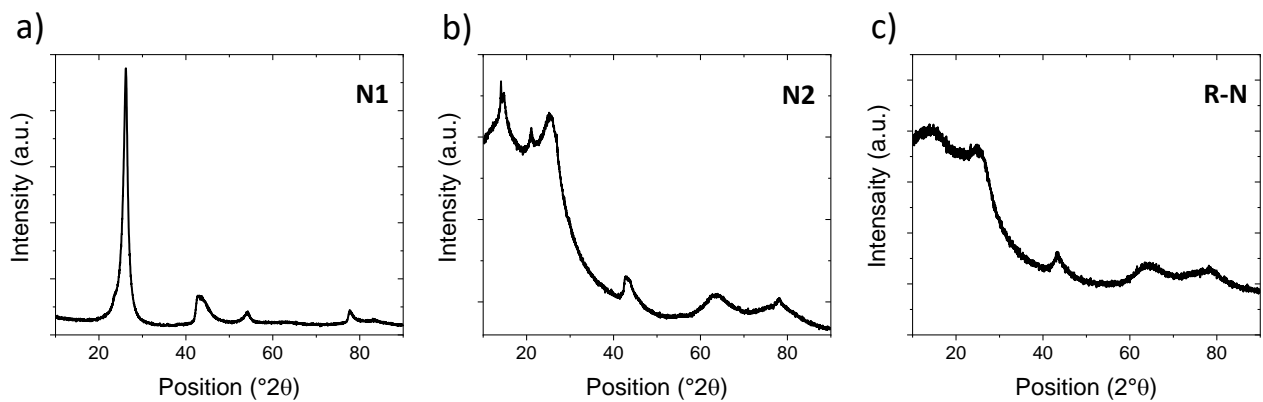

**Figure SI-3**. XRD diffractograms of a) N1; b) N2 and c) r-N.

Figure SI-3 presents the XRD patterns of the three N-graphenes investigated in the main text. The X-ray diffractograms of $N_2$ reduced (r-N) and nitrogen doped graphene (N1 and N2) show high disorder and are dominated by the (002) reflection corresponding to coherent and parallel stacking of graphene like-sheets ($2\theta \sim 25°$).[6]

When compared with graphite (d spacing of 0.3377 nm) all reduced and doped graphenes show higher interlayer spacing due to the presence of different functional groups, as residual oxygen functionalities from graphene oxide, and the high disorder induced by chemical reduction and nitrogen doping procedure.[7] The lowest d-spacing was observed for nitrogen doped graphene obtained from graphene oxide made by the Hoffman method (N1), showing a (002) reflection corresponding to d = 0.340 nm. In the nitrogen doped graphene prepared from graphene oxide made by the Tour method (N2) several diffraction maxima related to the (002) reflections can be observed, providing a broad spectrum with interlayer spacing ranging from 0.37 to 0.63 nm. Similarly, the $N_2$ reduced graphene oxide (r-N) shows a high degree of disorder indicated by significant broadening of the (002) reflection with two maxima appearing at 0.35 nm and 0.65nm. The weak diffraction pattern found at $2\theta = 43°$ originates from an in-plane diffraction pattern.

**b-** *Scanning Electron Microscopy (SEM) and X-ray Energy Dispersion Mapping (EDS)*

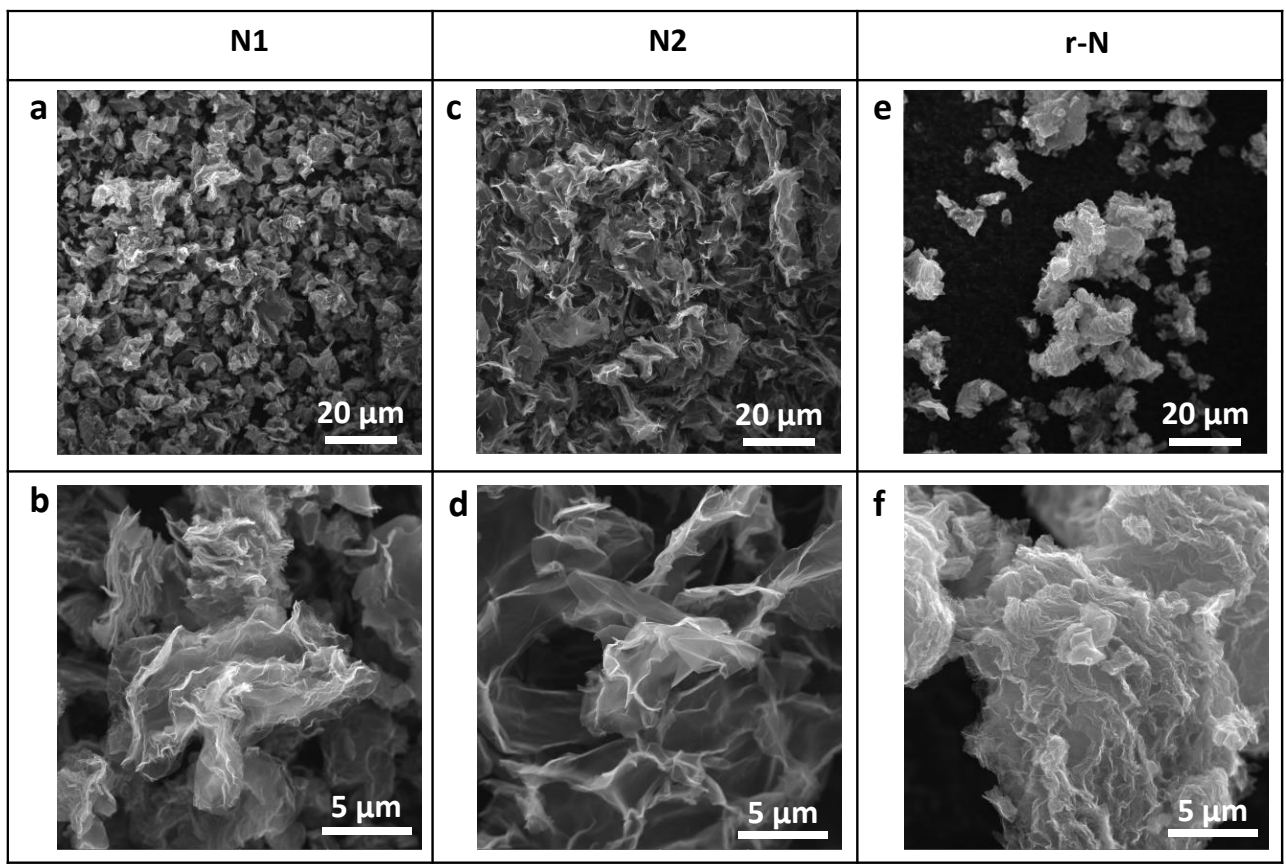

**Figure SI-4:** Scanning Electronic micrographs of the pristine N-graphene employed in this work: a- and b- N1-graphene; c- and d- N2-graphene; e- and f- r-N-graphene.

**Figure SI-4** shows the multilayer structure and micrometer size of the N-graphenes. Scanning electron microscopy (SEM) images are acquired using a Tescan Lyra dual-beam microscope with a FEG electron source with an accelerating voltage of 15 kV.

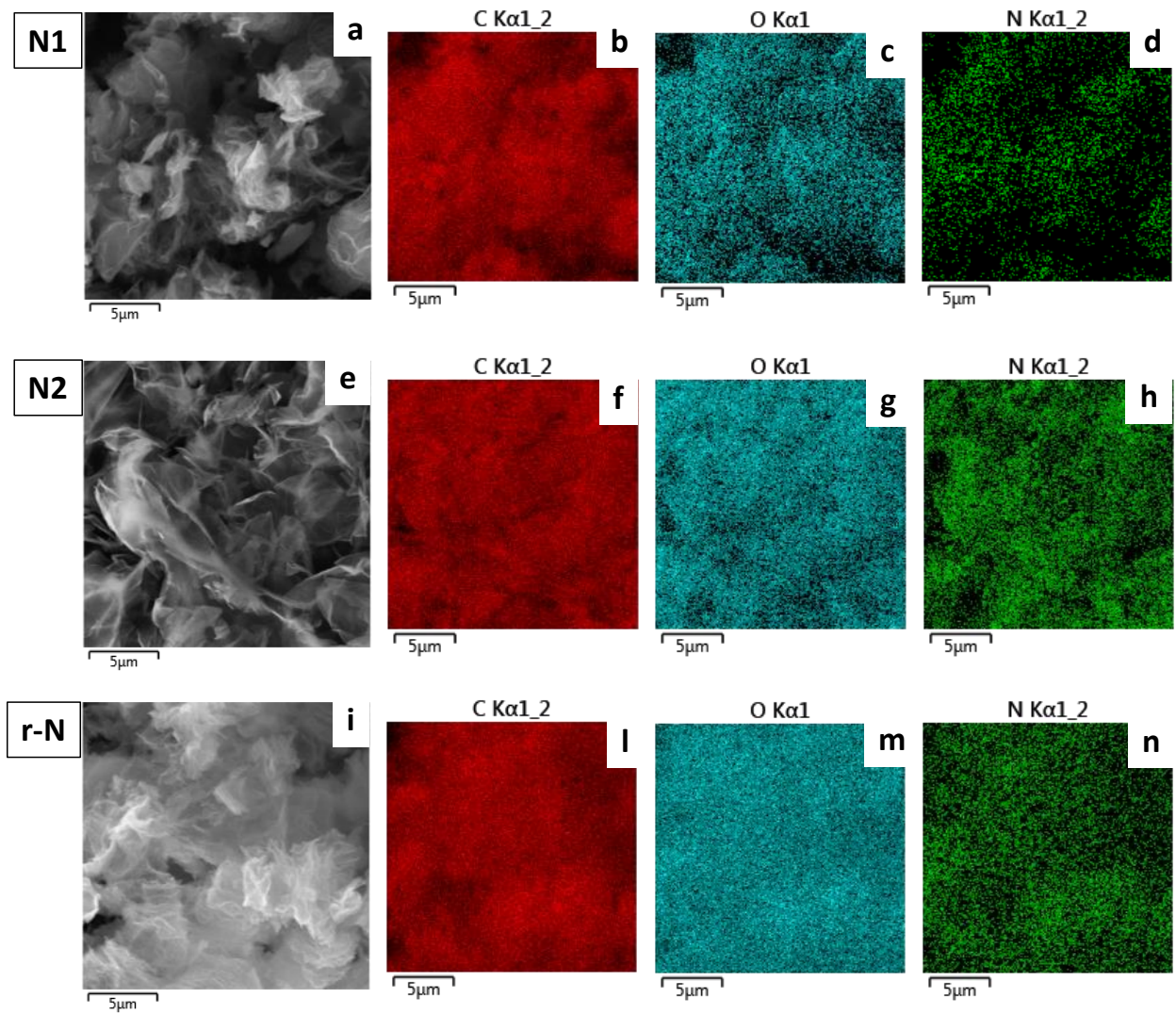

**Figure SI-5:** SEM micrograph (a- N1; e- N2; and i- R-N) and X-ray elemental maps of N-graphenes N1 (b. c. d); N2 (f, g, h); R-N (l, m, n).

EDS maps (Figure SI-5) show the homogeneous distribution of all elemental components over the N-graphenes samples, indicating the homogeneous N-doping.

Elemental composition and mapping were performed using an energy dispersive spectroscopy (EDS) analyzer (X-MaxN) with a 20 mm 2SDD detector (Oxford instruments) and AZtecEnergy software. The samples were placed on a carbon conductive tape to ensure the electrical discharge during measurement.[8]

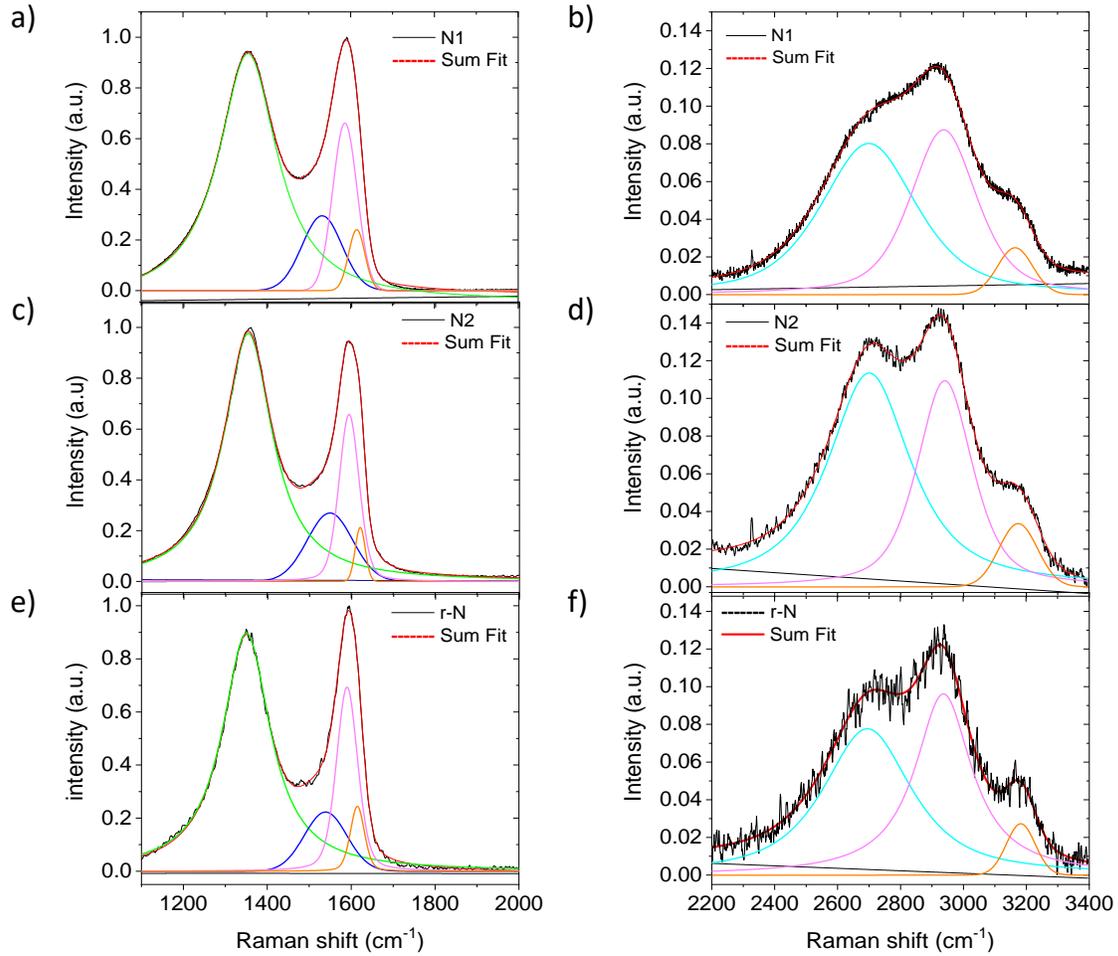

**Figure SI-6:** Selected spectra showing the fitting parameters for N1 (a-b); N2 (c-d); r-N (e-f).

**Figure SI-6** shows the peak deconvolution of the N1, N2 and r-N spectra. In the D and G peaks region (1000-2000 cm$^{-1}$) one additional peak is identified around 1550 cm$^{-1}$ and assigned to the D*, while the D´ is found between 1615 and 1620 cm$^{-1}$. In the 2000-3400 cm$^{-1}$ region we find three peaks assigned to 2D ($\sim$ 2700 cm$^{-1}$), D+D´ ($\sim$ 2940 cm$^{-1}$) and 2D´ ($\sim$ 3175 cm$^{-1}$).

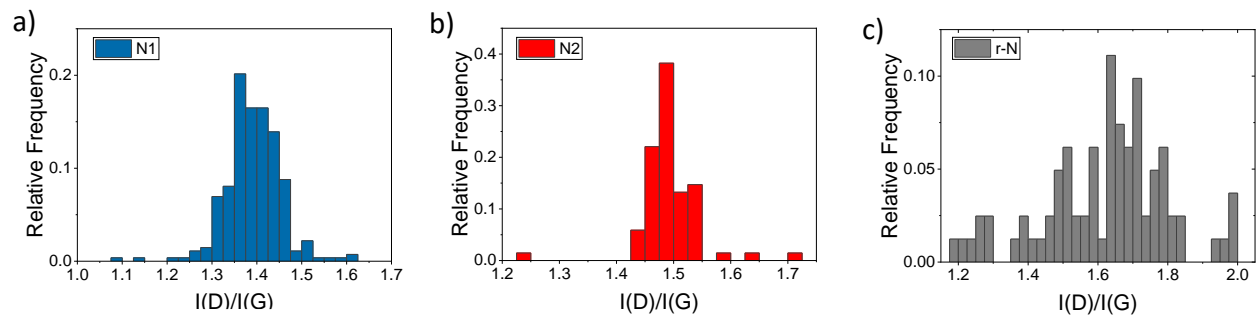

**Figure SI-7:** Distribution of the intensity ratio of the D and G peak, I(D)/I(G), per each of the investigated N-graphenes.

The high value found for I(D)/I(G) (> 1, **Figure SI-7**) confirms the presence of highly defective samples. The sharper distribution found for N2 demonstrates its higher homogeneity in defects distribution and electronic properties, compared to N1 and r-N.

### d- Ultraviolet Photoelectron Spectroscopy

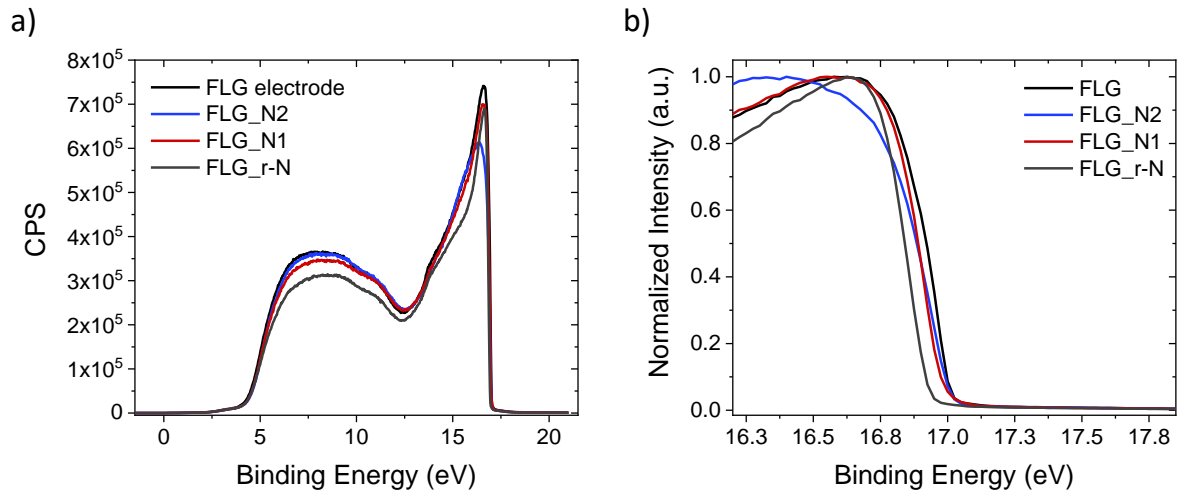

**Figure SI-8:** UPS data acquired on the few-layers graphene (FLG) based electrodes coated with different types of N-graphenes. a- full-scale UPS spectra. b- High binding energy region spectra showing the small difference in binding energy between the different electrodes, which falls within the energy resolution of our UPS (25 meV).

Ultraviolet photoelectron spectroscopy (UPS) with He I (hν = 21.21 eV) radiation was performed to estimate the Fermi energy level ($E_F$) and the valence band maximum of the materials under investigation (**Figure SI-8**).

## 2. Correlation between instant power and chemical and electrical properties of N-graphene

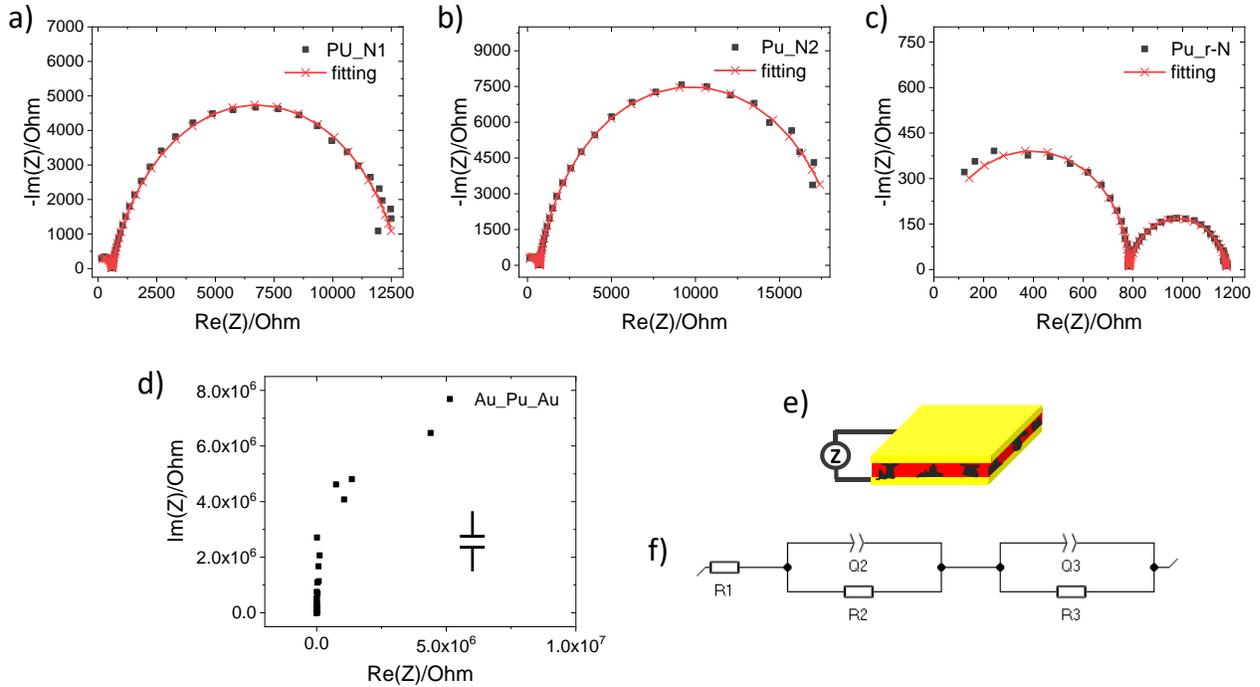

**Figure SI-9:** Impedance spectra and Z-fitting curve of devices: a- Au-PU_N1-Au; b- Au-PU_N2-Au; c- Au-PU_r-N-Au. d- Impedance spectra of device Au_PU_Au showing the purely capacitive response. e- schematic of the device structure used for impedance measurement. F- Equivalent circuit employed for the impedance spectra fitting of a- b- and c-.

The capacitance of the Au_Pu_Au device (**Figure SI-9d**) was calculated according to the impedance (Z) of a pure capacitor (C= 1/(jωZ), with ω being the frequency).

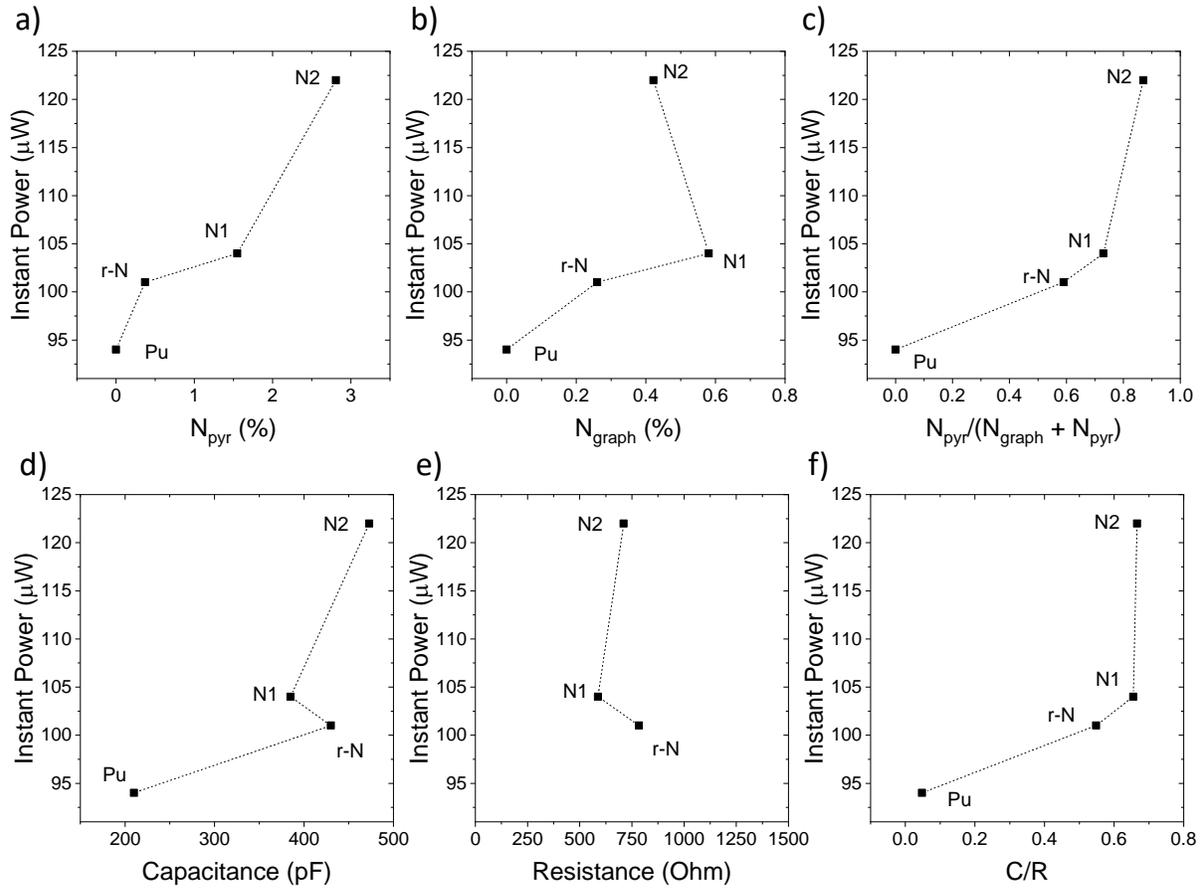

**Figure SI-10:** Dependence of the instant power on a) the content of N-pyridinic ($N_{pyr}$); b) the content of N-graphitic ($N_{graph}$); c) the relative content of $N_{pyr}$ and $N_{graph}$; d) the interlayer capacitance; e) the interlayer resistance; f) the ratio between capacitance (C) and resistance (R) of the interlayer.

**Figure SI-10** highlights the lack of correlation between the measured TENG's power output and the absolute value of either $N_{pyr}$, $N_{graph}$, C or R. A strong correlation is instead found for the TENG's power output and the relative ratio $N_{pyr}/(N_{pyr}+N_{graph})$ ratio and the C/R ratio.

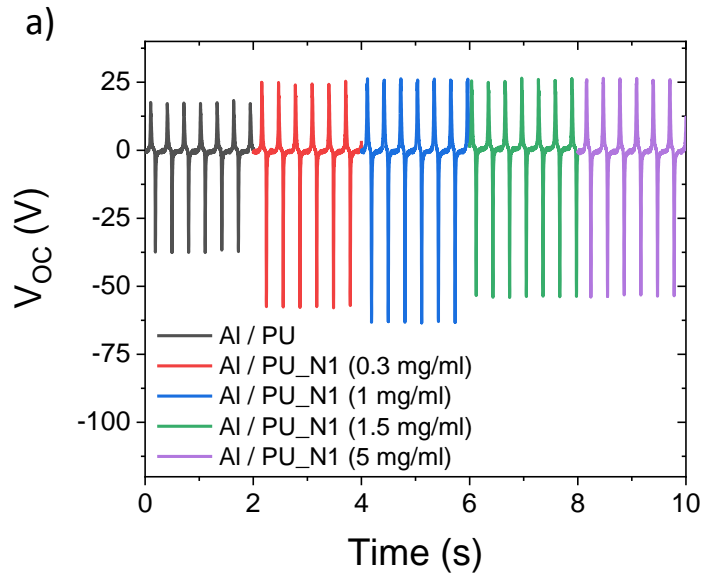

**Figure SI-11:** a) Open Circuit Voltage ($V_{OC}$) dependence upon the N-graphene concentration. $V_{OC}$ measured after 200 second of TENG cycling at 5 Hz, i.e., when no more significant variation of the $V_{OC}$ could be observed. (Applied force 10N, operational frequency 5 Hz, 4×4 cm$^2$, 5 mm; $V_{OC}$ acquired with 40 ΩM probe).

**Figure SI-11** show the $V_{OC}$ measured on TENGs comprising aluminium (Al) electrodes and nylon and PVDF as triboelectric layers. The N-graphene functionalized glue is introduced at the interface between the Al electrode and the nylon membrane. The N-graphene concentration reported in the figure refer to its concentration in isopropanol solvent before the addition to the PU glue.

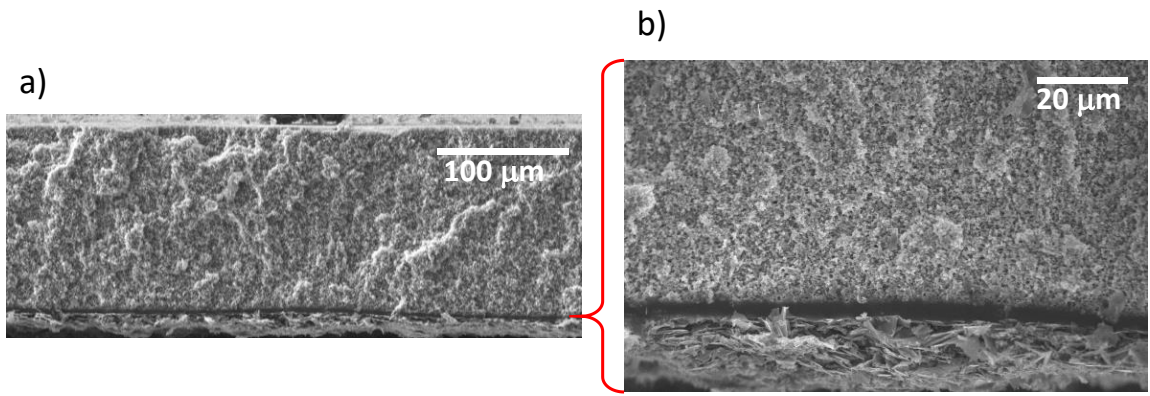

**Figure SI-12:** a) and b) SEM images showing the continuous and uniform N-graphene functionalized glue interlayer present at the interface between the FLG electrode and the PVDF triboelectric materials. b) magnification on selected area. Higher magnification image showing the presence of the N-graphene in the interlayer are reported in the main text.